\documentclass[twocolumn, aps, preprintnumbers, superscriptaddress, 
  prl]{revtex4-1}
\pdfoutput=1
\usepackage[pdftex]{graphicx}
\usepackage{amssymb,amsmath}
\usepackage[colorlinks,allcolors=blue,linktocpage]{hyperref}
\usepackage{natbib}
\usepackage{xcolor}

\begin{document}
\preprint{INR-TH-2023-006}
\title{Self-similar growth of Bose stars}

\author{A.S. Dmitriev}\email{dmitriev.as15@physics.msu.ru}
\affiliation{Institute for Nuclear Research of the Russian Academy
  of Sciences, Moscow 117312, Russia}
\author{D.G. Levkov}
\affiliation{Institute for Nuclear Research of the Russian Academy
  of Sciences, Moscow 117312, Russia}
\affiliation{Institute for Theoretical and Mathematical
  Physics, MSU, Moscow 119991, Russia}
\author{A.G. Panin}
\affiliation{Institute for Nuclear Research of the Russian Academy
  of Sciences, Moscow 117312, Russia}
\author{I.I. Tkachev}
\affiliation{Institute for Nuclear Research of the Russian Academy
  of Sciences, Moscow 117312, Russia}
  \affiliation{Novosibirsk State University, Novosibirsk 630090, Russia}

  \begin{abstract}
  We analytically solve the problem of Bose star growth in the bath of
  gravitationally interacting particles. We find that after
    nucleation of this  object the bath  is described by a
  self-similar solution of kinetic equation. Together with the
  conservation laws, this fixes mass evolution of the Bose
  star. Our theory explains, in particular, the slowdown of
  the star growth at a certain ``core-halo'' mass, but  also predicts
  formation of heavier and lighter objects in magistral dark matter
  models. The developed ``adiabatic'' approach to self-similarity may
  be of interest for kinetic theory  in general.
\end{abstract}

\maketitle

\paragraph{1. Introduction.}

Gravitationally bound blobs of Bose-Einstein
condensate~\cite{Ruffini:1969qy, *Tkachev:1986tr}~--- Bose stars~---
have regained a lot of attention recently. This is because they are
abundant in models with light dark matter~\cite{RingwaldPDG,
  *Niemeyer:2019aqm} consisting, e.g., of ``fuzzy'' bosons or QCD
axions. In those two cases, the Bose stars are
called ``solitonic galaxy cores''~\cite{Schive:2014dra} and ``axion
stars''~\cite{RingwaldPDG, *Niemeyer:2019aqm},
respectively. Typically, the self-couplings of light dark matter particles
are tiny and can be ignored. But their phase-space density is so
large~\cite{Tkachev:1991ka} that  thermalization can occur inside the
smallest cosmological structures via universal gravitational
interactions~\cite{Levkov:2018kau}. This makes the Bose star 
appear in the center of every such structure~\cite{Schive:2014dra,
  Levkov:2018kau, Eggemeier:2019jsu} in kinetic time. 

The question is, how do the newborn Bose stars grow? Lattice
simulations show that their masses  increase at first as
$M_{bs}\propto t^{1/2}$~\cite{Levkov:2018kau}  and then slow
down~\cite{Eggemeier:2019jsu}. But the numerical results on the
late-time behavior are conflicting: $M_{bs}\propto t^{1/8}$
in~\cite{Eggemeier:2019jsu, Chen:2020cef} and~$t^{1/4}$
in~\cite{Chan:2022bkz}.

In this Letter, we for the first time show~\footnote{Self-similar
solutions are well-known in kinetic theory with short-range
interactions~\cite{Semikoz:1994zp, *Micha:2002ey, *Micha:2004bv, 
  *SEMISALOV2021105903} and in dynamical long-range
problems like collapse~\cite{Choptuik:1992jv, *Maeda:2004kw,
  *Gundlach:2007gc} or infall~\cite{Bertschinger:1985pd, *Sikivie:1996nn}. But their
relevance for kinetics caused by gravitational (long-range) scattering
was not observed before.} that Bose-Einstein condensation of dark
  matter via gravitational (long-range) scattering is described by 
self-similar solutions of kinetic 
equation. Computing  the condensation flux onto the Bose star, we
analytically obtain its growth law. The star  mass is not a simple
power of time, but can be piecewise approximated by all of the
behaviors above.

\paragraph{2. A crucial observation.}
\label{sec:2.-cruc-observ}
\begin{figure}
  \centering
  \includegraphics{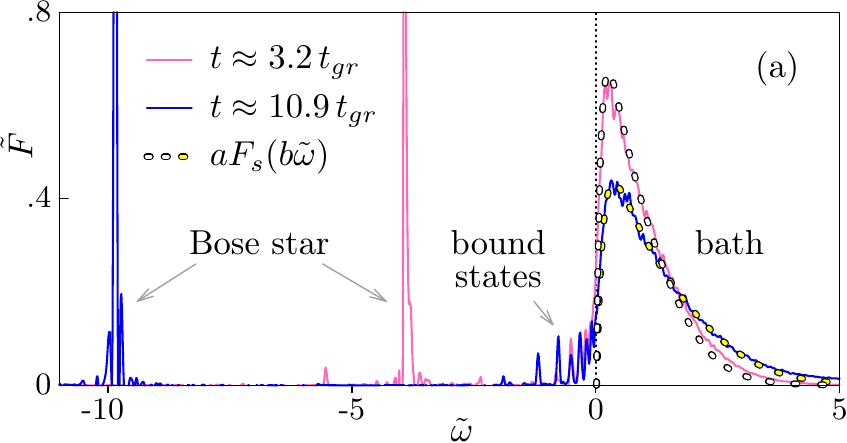}

  \vspace{2mm}
  \includegraphics{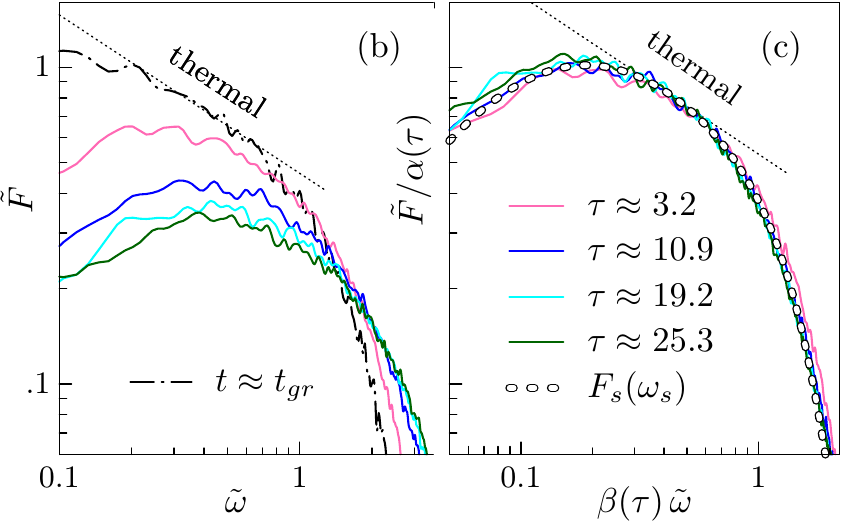}
  \caption{Simulation of Eq.~\eqref{eq:GP2} with $M = 50 \, p_0/m^2G$ and $L =
    60/p_0$.  (a)~Spectra~\eqref{eq:1} at two moments of time (solid
    lines). (b)~Bath spectra (${\omega>0}$) 
    at large times and (c)~their self-similar
    transformation~\eqref{eq:ab} with~${D = 
      2.8}$. Figure (b)  includes $t\approx t_{gr}$ graph
    (dash-dotted). Chain points in Figs.~(a), (c) show 
    the solution of Eq.~\eqref{eq:4} with $D=2.8$ and the source;
    see Supplemental Material~C (SM-C) for parameters.}
  \label{fig:spectra}
\end{figure}

Consider a cloud of nonrelativistic dark matter bosons inside the smallest
cosmological structure: a minicluster of axions~\cite{Kolb:1993zz,
  *Kolb:1993hw, *Vaquero:2018tib, *Buschmann:2019icd,
  *Eggemeier:2019khm, *Ellis:2020gtq} or a galaxy halo of ``fuzzy''
dark matter. At small particle masses $m$ the occupation numbers are
so large that the bosons are
described by a random classical field $\psi(t,\, \boldsymbol{x})$
evolving in its own gravitational potential~$U(t,\,  \boldsymbol{x})$,
\begin{align}
  \label{eq:GP2}
  &i \partial_t \psi = -\Delta \psi/2m  + mU \psi \,, \\
  \notag
   &\Delta U = 4\pi G\,  (m|\psi|^2 - \bar{\rho}) \,,
\end{align}
where $\bar{\rho} \equiv M/L^3$ is the mean density and $M$ is the
total mass. For simplicity, we  approximate the
structure with periodic box of size $L$. 

We solve Eqs.~(\ref{eq:GP2}) using a stable 3D code of
Ref.~\cite{Levkov:2018kau}. The starting point of this evolution is a
virial equilibrium, i.e. Gaussian-distributed field with
Fourier image $|\psi_{\boldsymbol{p}}|^2 \propto M 
\mathrm{e}^{-\boldsymbol{p}^2/p_0^2}$ and random
phases~$\arg\psi_{\boldsymbol{p}}$; here $\omega_0\equiv p_0^2/2m$ 
is the typical particle energy~\footnote{Equations~\eqref{eq:GP2}
  have exact scaling symmetry changing~$p_0$; see,
  e.g.,~\cite{Schive:2014dra, Levkov:2018kau}. This makes the solution
  depend on dimensionless combinations~$p_0 \boldsymbol{x}$, 
  $\boldsymbol{p}/p_0$, and $\omega / 2\omega_0$.}. We study 
mass distribution    
$F(t,\,\omega) = dM/d\omega$ of bosons over energies $\omega$. It is
given by time Fourier transform~\cite{Levkov:2018kau}:  
\begin{equation}
  \label{eq:1}
  F = m \int \frac{dt'}{2\pi} \, d^3\boldsymbol{x} \, \psi^*(t,
  \boldsymbol{x}) \psi(t+t',\boldsymbol{x})\,  \mathrm{e}^{i\omega 
  t' - t'^2/\Delta t^2 }\,,
\end{equation}
where $\Delta t^{-1} \ll \omega_0$ is the energy resolution. In the
isotropic homogeneous case, this function is related to the usual
phase-space density $f(p)$ as $F =L^3m^2pf/2\pi^2$ with $\omega =
p^2 /2m$. Below we exploit dimensionless units:
$\tilde{F}(t,\, \tilde{\omega}) \equiv 2\omega_0 F/M$ and $\tilde{\omega}
\equiv \omega/2\omega_0$.

Our ensemble has large occupation numbers and hence thermalizes into a
Bose-Einstein condensate. This is seen in
simulations~\cite{Levkov:2018kau} as phase transition at a
  kinetic time $t = t_{gr} \equiv 2b\sqrt{2}\,
  \omega_0^3/ [ 3\pi^3 {G^2 \bar{\rho}^2  \ln(p_0L)}]$, where  ${b
    \approx 0.9}$ for the Gaussian initial 
distribution. Namely, after $t_{gr}$ the spectrum $F(t,\,\omega)$
develops a narrow peak at $\omega \approx \omega_{bs}<0$ moving with
time to lower energies; see Fig.~\ref{fig:spectra}(a) and the  
video~\cite{movie}. The peak is a Bose star~\cite{Ruffini:1969qy,
    *Tkachev:1986tr, RingwaldPDG, *Niemeyer:2019aqm}: a condensate of
particles occupying a single~--- ground~--- level~$\omega_{bs}$ in 
the  collective gravitational well~$U$.  Once the  
Bose  star appears, the ensemble  mass  $M = M_{bs}
+ M_{e} + M_{b}$ divides between this  object ($M_{bs}$), excited
bound states in its gravitational field ($M_{e}$), and the ``bath'' of 
particles with $\omega>0$~($M_{b}$).  The conditions for
condensation are still satisfied, so $M_{*}$ grows at
$t>t_{gr}$.

Below we measure time in kinetic intervals $\tau \equiv
t/t_{gr}$ and compute $M_i$ by
integrating  $F(\omega)$ over the respective regions. E.g., the
``dressed star'' mass is $M_* \equiv M_{bs} + M_{e} = \int_{\omega<0}
F\, d\omega$,  while $M_b$ and $M_{bs}$ are the integrals
over $\omega>0$ and $\omega \approx \omega_{bs}$,
respectively.

Now, we make an important observation. Consider the  $\omega>0$
spectrum,  i.e.\ the ``bath''. It changes a lot after the 
Bose star formation,  cf.\  the graphs with different~$\tau$ in
Fig.~\ref{fig:spectra}(b). The same graphs, however, coincide
in Fig.~\ref{fig:spectra}(c) after time-dependent rescaling of~$F$
and~$\omega$:
\begin{equation} 
  \label{eq:ab}
  \tilde{F}(t, \,\tilde{\omega}) = \alpha F_s(\beta\tilde{\omega})\,,
  \quad
 \alpha = \tau^{-1/D}\,, \quad    \beta = \tau^{2/D -1}\,,
\end{equation}
where $D = 2.8$. This means that the bath is self-similar and can
be fully described by a function $F_s(\omega_s)$. Below we
demonstrate that Eq.~\eqref{eq:ab} is an attractor solution:  
kinetic evolution generically approaches it at large~$t$.

It is worth noting that Bose star formation can be perceived as  a
second-order critical phenomenon. First, its order parameter
  $M_{bs}(t)$ grows from zero at ${t\geq t_{gr}}$. Second, the
  bath spectrum  has thermal small-$\omega$ tail ${F \propto
    \omega^{-1/2}}$ at $t\approx t_{gr}$, see
  Fig.~\ref{fig:spectra}(b).  In Supplemental Material~A (SM-A) we
  show that this entails power-law field correlators at large
  distances. The thermal parts remain in the self-similar spectra at
$t>t_{gr}$, cf.\ Fig.~\ref{fig:spectra}(c).

\paragraph{3. Self-similar attractor.}
Let us ignore the effect of Bose star gravitational field on
the bath. Then evolution of $F(t,\,\omega)$ at $\omega>0$ is
governed  by  a homogeneous and isotropic kinetic
equation~\cite{pitaevskii2012physical, *2013PhyU...56...49Z,
  *Skipp:2020xcc, Levkov:2018kau}
\begin{equation}
    \label{eq:2}
  \partial_\tau \tilde{F} = \mathrm{St}\, \tilde{F}\,,
\end{equation}
where $\mathrm{St}\, \tilde{F}$ is the Landau scattering integral~---
functional of $\tilde{F}(\tilde{\omega})$ at given~$\tau$,
see its explicit form in~\cite{Levkov:2018kau} and~SM-B.

Dramatically, the ansatz~(\ref{eq:ab}) 
passes through Eq.~(\ref{eq:2}) at any~$D$ leaving a
one-dimensional equation for the profile,
\begin{equation}
  \label{eq:4}
(2/D-1) \, (\omega_s
  \partial_{\omega_s}F_s) - F_s/D = \mathrm{St}\,  F_s\,,
\end{equation}
  This is
guaranteed by the scaling ${\mathrm{St}\, \tilde{F} =\alpha^3
  \beta\, \mathrm{St}\, F_s}$ reflecting long-range nature of
gravitational scattering,  see~\cite{Levkov:2018kau}
  and~SM-B. The scaling  is generic: one 
can find it even using the estimate~$\mathrm{St}\, F
  \sim {F/t_{gr}}$.  
  
On the other hand, Eq.~(\ref{eq:ab}) is {\it not} a solution if
  the bath is isolated. Indeed, self-similarity gives
time-dependent mass ${M_b  \propto    \tau^{k_M}}$ and energy ${E_b  
  \propto  \tau^{k_E}}$ with
\begin{equation}
 \label{eq:k}
 k_M = 1 - 3/D\, , \quad    k_E =  2-5/D\,,
 \quad 3 k_E - 5k_M = 1\,.
\end{equation}
This contradicts to the conservation laws.

But the ongoing condensation radically changes the boundary conditions
for the bath. Indeed, the bath bosons may scatter, loose energy, and
append either to the Bose star or to one of 
its bound states  at $\omega<0$. Besides, with time
the star gravitational well grows deeper and adiabatically drags
low-energy particles to $\omega<0$. Both mechanisms absorb bosons with 
$\omega \approx   0$, since gravitational scattering is more effective
at low transfers $\Delta \omega \ll
\omega_0$~\cite{pitaevskii2012physical, *2013PhyU...56...49Z}. This
heats the remaining ensemble  due to energy conservation. As a result,
the bath has decreasing~$M_b$ and growing  $E_b$, i.e.\ $5/2 < D  < 3$
in Eq.~(\ref{eq:k}).

To account for condensation in Eq.~\eqref{eq:4}, we impose
  a condition of finite particle flux at ${\omega_s
    \approx 0}$ and add an energy source $\mathrm{St}\, F_s \to  
\mathrm{St}\, F_s + J_s(\omega_s)$ to the right-hand side. This
  gives a family of solutions at $D \geq   5/2$; see SM-C for
  details. The solution  $F_s(\omega_s)$ with $D=2.8$ and properly
selected $J_s(\omega_s)$ is shown in Figs.~\ref{fig:spectra}(a), (c) by
chain points. Having almost constant condensation flux at low~$\omega_s$,
it nevertheless considerably differs from the power-law Kolmogorov 
  cascades~\cite{zakharov2012kolmogorov}.

It is crucial that the self-similar solutions~(\ref{eq:ab}) are
attractors of kinetic evolution. This property is apparent in 
Fig.~\ref{fig:spectra}(c), but we confirm it explicitly in
  SM-D by solving the full kinetic equation~\eqref{eq:2} with
time-dependent source $\tilde{J}(\tau,\, \tilde{\omega})$. Even
if $\tilde{J}$ is essentially non self-similar, the solution
$\tilde{F}(\tau,\, \tilde{\omega})$ approaches Eq.~\eqref{eq:ab}
with some $D$.


\paragraph{4. Growth of the Bose star.}
In our problem, self-similarity of the bath is broken by   the Bose
star which injects energy at its own, non scale-invariant rate $J$. On
the other hand, the self-similar solutions are attractors. This
implies an ``adiabatic'' regime which was never studied before:
the bath remains almost self-similar at all times, but its
parameters  slowly drift with time.

\begin{figure}
   \centerline{\includegraphics{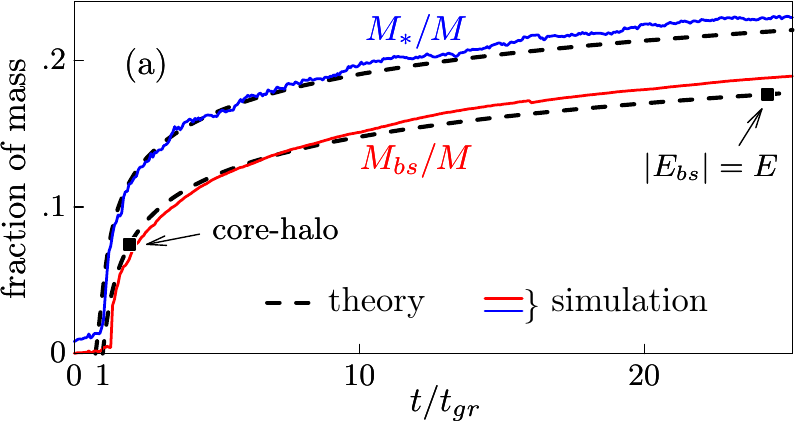}}

   \vspace{1mm}
   \centerline{\includegraphics{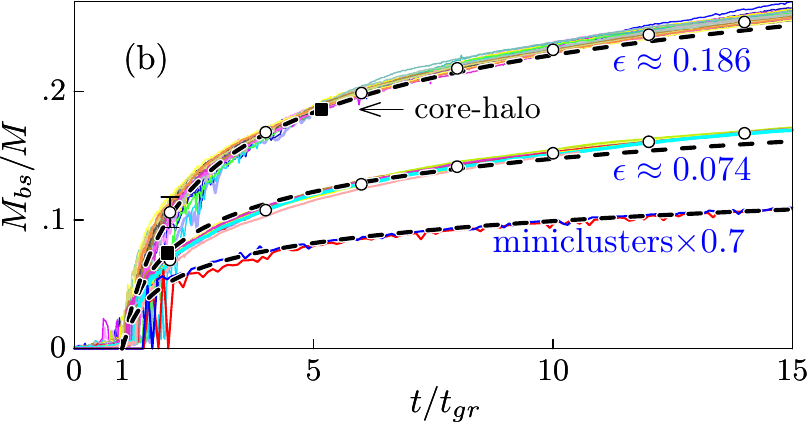}}
   \caption{(a) Bose star mass $M_{bs}(t)$ and the mass of the ``dressed''
     object $M_*(t)$ in the box simulation of Fig.~\ref{fig:spectra} with
     $\epsilon\approx 0.074$. (b)~Evolutions of $M_{bs}(t)$ in 
     $11+22$ box simulations with essentially different $t_{gr}$
     at two values of  $\epsilon$ (two upper graphs). Numerical results are
     shown  by solid lines, while dashed is the theory~(\ref{eq:3}) 
     with $x_e \approx 0.043$ and $0.021$ at $\epsilon \approx 0.074$
     and $0.186$, respectively. Circles average over simulations with
     given~$\epsilon$. The lower graph shows 2 minicluster
       simulations at $\epsilon \approx 0.066$ (solid lines) fitted by the theory
       with $x_e\approx 0.026$ and $\tau_i = -0.1$ (dashed). For
       visualization purposes, we rescaled the minicluster lines by
       $M_{bs} \to 0.7 M_{bs}$. 
   }
  \label{fig:Mbs}
\end{figure}

  In the first~--- crude~---  
  approximation we can account for time dependence of $D =
  D(\tau)$. Define $k_M(\tau) \equiv  d\ln M_b/d\ln \tau$ 
and $k_E(\tau) \equiv d\ln E_b/d\ln \tau$. We assume that they
satisfy the self-similar law~(\ref{eq:k}), ${3k_E -
  5k_M \approx 1}$, if they change slowly. Then the conservation laws
$M_b =  M - M_*$ and  
$E_b = E - E_*$ give $d\!\ln \tau \approx 3 d\! \ln(E-E_*) - 5d\!\ln
  (M-M_*)$ or, integrating,
\begin{equation}
 \label{eq:result}
 (1- E_*/E)^3(1-M_*/M)^{-5} \approx (\tau - \tau_i)/\tau_*\,,
\end{equation}
where $\tau_*$ is an integration constant,  $M_{*}$ and $E_*$ are
  the parameters of the ``dressed'' star, and we recalled the
time translations $\tau \to \tau - \tau_i$~\footnote{Our best-fit
value $\tau_i = -0.1$  from Figs.~\ref{fig:Mbs} and~SM-S2 is
quite small and does not affect the  agreement in
Fig.~\ref{fig:spectra}(c).}.
 
To extract the Bose star mass evolution from Eq.~(\ref{eq:result}), we
estimate the contributions of the excited discrete levels at
  $\omega<0$. Theory suggests that large-mass condensate cannot be
accumulated on those levels: it would become unstable once
gravitationally  self-bound~\cite{Lee:1988av,
  Dmitriev:2021utv}. Then $M_e < M_{bs}$. This is confirmed by our 
  simulations: $ M_e(t) \equiv M_* - M_{bs} $  is small and almost constant in
Fig.~\ref{fig:Mbs}(a) at $\tau \gtrsim 2$. Moreover, the excited
  levels with $\omega<0$ carry negligibly small energy in simulations
  as compared to the Bose star itself:
$E_* \approx {E_{bs} =  -\gamma M_{bs}^3}$,  where $\gamma \approx
0.0542\,  m^2 G^2 $~\cite{Dmitriev:2021utv}. Indeed, in
Fig.~\ref{fig:spectra}(a) only the bound states with $\omega\approx 0$
are occupied. Taking $E_* \approx E_{bs}$~\footnote{More
precise expression follows from the adiabatic  theorem: $E_e = - \zeta
M_{bs}^2$,  where $\zeta$ depends on the occupation
numbers of the bound states.} and constant $x_e \equiv M_{e}/M$, we
obtain the growth law for $x_{bs} (\tau) \equiv M_{bs}/M$:   
\begin{equation}
  \label{eq:3}
  (1 + x_{bs}^3/\epsilon^2)^3 (1 - x_e
  - x_{bs})^{-5} \approx (\tau - \tau_i) / \tau_*\,.
\end{equation}
Here $\epsilon^2 \equiv E/\gamma M^3$ is a combination of the
  total mass and energy proportional to
  the invariant $\Xi$ from Refs.~\cite{Schwabe:2016rze, Mocz:2017wlg}, 
  while  $M_*(\tau) = (x_{bs}+x_e) M$. 

  Note that $\tau_i$, $\tau_*$, and $x_e$   in Eq.~(\ref{eq:3})
are empiric fitting parameters. However, $\tau_* \approx
(1-\tau_i)(1 - x_e)^5$ is fixed by the initial condition $M_{bs}
= 0$ at $\tau=1$, while ~$x_e$ is small and can be  ignored, if
unknown. This leaves only $\tau_i$ to fit; in fact, $\tau_i
  \approx -0.1$ agrees with all simulations in Fig.~\ref{fig:Mbs}.

In Fig.~\ref{fig:Mbs}(a) we show that the theory~(\ref{eq:3}) (dashed
lines) reproduces the simulation results for $M_{bs}(\tau)$ and
$M_*(\tau)$ (solid).  A significant statistical
test is shown in Fig.~\ref{fig:Mbs}(b) where we display
$M_{bs}(\tau)$ for 11 simulations with $\epsilon \approx 0.074$
and~$22$ simulations with $\epsilon \approx  0.186$ (solid 
data vs.\   dashed theory). These runs have
essentially different    parameters and kinetic times
$10^{3} \lesssim    \omega_0 t_{gr} \lesssim  3\cdot
10^{4}$. Nevertheless, their   graphs in   Fig.~\ref{fig:Mbs}(b) merge
into two distinct curves   at two values of $\epsilon$, which agree
with Eq.~(\ref{eq:3}).
Another strong test is performed in SM-E by considering
  self-interacting bosons. In this case our theory still describes
  numerical data, although Eq.~(\ref{eq:3}) gets modified
  by the Bose star self-interaction energy.

For gravitationally self-bound bath,  $p_0L\sim 5/\epsilon$. This
  means that
  kinetic approach is valid at $\epsilon \ll 1$~\cite{Levkov:2018kau}. Present-day 
  simulations~\cite{Kolb:1993zz, *Kolb:1993hw, 
    Schive:2014dra,Vaquero:2018tib, Buschmann:2019icd,
    Eggemeier:2019khm, Ellis:2020gtq} are restricted to 
  $\epsilon \gtrsim 0.05$, cf.\ Fig.~\ref{fig:Mbs}.  At these values,
  the Bose star growth is ``adiabatic'' from the start: $d
  k_{M,E}/d\ln(\tau   - \tau_i)  < {0.03/\epsilon < 1}$. At smaller
  $\epsilon$, adiabaticity is met at later stages. 

\paragraph{5. Core-halo relation and beyond.}
At $t\approx   t_{gr}$ the energy of the baby Bose star
is negligible, since ${E_{bs}\propto M_{bs}^3}$. Then
Eq.~(\ref{eq:result}) gives $M_*(t) \approx
M\, (\tau  - 1)/5\tau_*$~--- 
linear growth law  for the ``dressed'' object.

During longer initial stage, the star is still small,
${E_{bs} \ll E}$, and  Eq.~(\ref{eq:3}) 
linearizes to $3x_{bs}^3/\epsilon^2  + 5 x_{bs} \approx  (\tau -
1)/\tau_*$. Hence, at $x_{bs}\gtrsim\epsilon$ the 
 evolution slows down to ${M_{bs} \propto  t^{1/3}}$. This
  transition happens at $M_{bs} = \epsilon M$ when the virial
  velocities of the Bose star and the bath equalize, $|E_{bs}/M_{bs}| = E  
/M$, i.e.\ precisely at the ``core-halo'' point of
Refs.~\cite{Schive:2014hza, Bar:2018acw}. The time to the
slowdown is short at
small $\epsilon$:  $t   -t_{gr} \sim 9\epsilon \,t_{gr}$. This explains, why 
the  stars with $M_{bs} \sim \epsilon M$ form in cosmological 
simulations~\cite{Schive:2014dra, Schive:2014hza} and seemingly
do not grow any further.  In truth, the growth
continues~--- hence scatter~\cite{Schwabe:2016rze, Mina:2020eik,
  *Zagorac:2022xic, Chan:2021bja, *Nori:2020jzx} in the simulation
results for~$M_{bs}$.

The next slowdown  in Eq.~(\ref{eq:3}) occurs at
$E_{bs}= E$ and $M_{bs} =\epsilon^{2/3} M$. Such heavy objects
were observed in some cosmological simulations~\cite{Mocz:2017wlg,
  Mina:2020eik,  *Zagorac:2022xic}. After this point,
${M_{bs} \propto t^{1/9}}$.  Together, our laws $M_{bs} \propto
t^{1/3}$ and $t^{1/9}$ agree with numerical
data in ~\cite{Chan:2022bkz} and~\cite{Eggemeier:2019jsu,Chen:2020cef}.   

\paragraph{6. Bose star growth in a halo.}
Now, consider a denser gas which quickly forms a gravitationally bound
   halo/ minicluster under Jeans instability~\cite{Levkov:2018kau, Schwabe:2016rze, 
    Chen:2020cef}. Using the distribution~$F$ at $\omega < 
  0$, we find the minicluster mass $M$, its virial energy 
  $E_{\mathrm{mc}} < 0$, and mean particle energy $\omega_0  =
  -mE_{\mathrm{mc}}/M$. We also  compute its central density $\bar{\rho}
  = \rho(0)$ and potential $U(0)$. This gives~$t_{gr}$, the
  energy $E = E_{\mathrm{mc}} - U(0)M {> 0}$ counted from the lowest 
  level inside the halo, and $\epsilon^2 \equiv E / \gamma
  M^3$; see details in SM-F.

  With time, the minicluster gives birth to a Bose star. The growing
  mass of the latter is shown in Fig.~\ref{fig:Mbs}(b) for two
  simulations with $\epsilon  \approx 0.066$ (thin solid
  lines). Notably, $M_{bs}(t)$ is still described by the
    self-similar theory  with ${\tau_i = -0.1}$ (dashed line), where
  $x_e$ is   
  extracted from~$F$. This coincidence strongly supports our
  theory, as it occurs despite the fact that Eq.~(\ref{eq:3}) ignores
  inhomogeneity of the minicluster.

  \paragraph{7. Discussion.}
In this Letter we demonstrated that kinetics of Bose-Einstein
condensation is self-similar if it is governed by 
  gravitational (long-range) scattering. This solves a
  long-standing problem~\cite{2013PhyU...56...49Z,
    zakharov2012kolmogorov, Skipp:2020xcc} with absence of Kolmogorov power-law
  cascades in such systems. The Bose star growth
  law~(\ref{eq:result}),~(\ref{eq:3}) was derived using the new
  working assumption on the ``adiabaticity''  of scaling
  exponents. This framework may be useful in other contexts.

To date, simulations of light dark matter structure
  formation~\cite{Schive:2014dra, Schwabe:2021jne, Li:2020qva} cannot
  provide global distribution of Bose stars which are just too
  small. For that, one needs a theoretical input from our
  Eq.~(\ref{eq:3}) and Refs.~\cite{Kolb:1993zz, *Kolb:1993hw,
    *Vaquero:2018tib, 
    *Buschmann:2019icd, *Eggemeier:2019khm, *Ellis:2020gtq},
  cf.~\cite{Eggemeier:2021smj, *Ellis:2022grh, *Du:2023jxh}. 
Consider, e.g.,  growing Bose (axion) stars inside QCD axion
miniclusters~\cite{Kolb:1993zz, *Kolb:1993hw, *Vaquero:2018tib,
  *Buschmann:2019icd,  *Eggemeier:2019khm, *Ellis:2020gtq}. The latter
originate from the axion
overdensities $\Phi=\delta\rho/\rho|_{\mathrm{RD}} \lesssim 10$  at
the radiation-dominated epoch. Equation~(\ref{eq:3}) tells us that the star 
eats the fraction $x_{bs} \sim 0.1$ of the host minicluster in time $t
\sim t_{gr} x_{bs}^9/\epsilon^6$. This time is shorter than the the
age of the Universe if $\Phi \gtrsim 6 \, (10\, x_{bs} m_4)^{9/8}
M_{13}^{3/4}$, where we used the estimates of~\cite{Kolb:1994fi,
  Levkov:2018kau} and normalized $M_{13} \equiv M/(10^{-13}\,
M_{\odot})$ and $m_4 \equiv m/(10^{-4} \, \mbox{eV})$ to the centers
of the discussed minicluster and QCD axion mass
windows~\cite{Kolb:1993zz, *Kolb:1993hw,   *Vaquero:2018tib, 
  *Buschmann:2019icd, *Eggemeier:2019khm, *Ellis:2020gtq,
  Klaer:2017ond, *Gorghetto:2018myk, *Gorghetto:2020qws}. It is
thus realistic to expect that large parts of the densest 
miniclusters are nowadays engulfed by their axion
stars. Note that the latter may lead to spectacular
  observational effects, see e.g.~\cite{Levkov:2020txo, *Eby:2021ece,
    *Visinelli:2021uve, *Escudero:2023vgv}.

The other popular model describes growth of
gigantic Bose stars inside ``fuzzy'' dark matter galaxy
halos~\cite{Schive:2014dra}. Such stars do not
reach the ``core-halo'' point if the required time $\Delta t \sim  9\epsilon
t_{gr}$ exceeds the age of the Universe. This happens if 
$m \gtrsim 6\cdot 10^{-21}\, \mbox{eV} \cdot v_{30}^{5/2} \,
M_8^{-3/2}$, where we normalized the  virial velocity $v_{30} \equiv
v/(30 \ \mbox{km}/\mbox{s})$ and mass $M_8 \equiv M / (10^8\,
M_{\odot})$ to  the smallest dwarf galaxies. We see that the ``fuzzy''
Bose stars should be undergrown in all galaxies, if the current
experimental bound $m \gtrsim 2\cdot 10^{-20}\,
\mbox{eV}$~\cite{Rogers:2020ltq} on the  particle mass is satisfied.   

This Letter is dedicated to the memory of Valery Rubakov and
  Vladimir Zakharov. We thank J.~Chan, J.~Niemeyer, X.~Redondo, and
S.~Sibiryakov for discussions. The work was supported by the grant RSF
22-12-00215 and, in its numerical part, by the ``BASIS'' foundation.

\onecolumngrid

\vspace{5mm}
\hrule

\vspace{10mm}

\begin{center}
\textbf{\large {\it Supplemental material on the article:}\\[.2em]
  Self-similar growth of Bose stars}
\end{center}

\vspace{5mm}

\twocolumngrid
\setcounter{equation}{0}
\setcounter{figure}{0}
\setcounter{table}{0}
\makeatletter
\renewcommand{\theequation}{S\arabic{equation}}
\renewcommand{\theHequation}{S\arabic{equation}}
\renewcommand{\thefigure}{S\arabic{figure}}
\renewcommand{\theHfigure}{S\arabic{figure}}
\renewcommand{\thetable}{S\arabic{table}}
\renewcommand{\theHtable}{S\arabic{table}}

\section*{A. Distribution function}
\label{sec:s1.-distr-funct}
In the weakly coupled gas, the field~$\psi(t,\, \boldsymbol{x})$
evolves almost freely in the mean gravitational field which, in
turn, changes slowly due to rare scatterings. This means that
at timescales ${\Delta  t \ll t_{gr}}$ we can
write~${U \approx \langle U(\boldsymbol{x})\rangle}$ and 
\begin{equation}  \label{eq:s2} 
  \psi(t,\, \boldsymbol{x}) \approx \sum_{n} f_n \,
  \psi_n(\boldsymbol{x}) \, \mathrm{e}^{- i\omega_n t}\,.
\end{equation}
Here $\{ \psi_n,\, \omega_n\}$  is the instantaneous eigenspectrum
of~$\langle U\rangle$ and $|f_n|^2$ are the occupation
numbers of levels~$\omega_n$. Substituting~(\ref{eq:s2}) into
Eq.~\eqref{eq:1}, we get,
$$F \approx m\sum_n 
  |f_n|^2 \,{ \tilde{\delta}(\omega - \omega_n)}\,,$$ where
$\tilde{\delta}(x) = \mathrm{e}^{-(x\Delta   t/2)^2} \Delta
t/\sqrt{4\pi}$  is the smoothed $\delta$-function. This confirms that
Eq.~\eqref{eq:1} defines the distribution function
$F\approx dM/d\omega$, indeed. Resolution~$\Delta \omega \sim
    (\Delta t)^{-1}$ of the latter is of order~$t_{gr}^{-1} \ll
  \Delta \omega \ll \omega_0$. 

In Sec.~2 of the main text we mention that the distribution
  of particles in the box acquires  thermal
  low-$\omega$ tail ${F \propto\omega^{-1/2}}$ at $t\approx
    t_{gr}$. Let us show that  this 
  entails power-law correlator of the field at large
  distances. Consider the bath of unbound particles in the
periodic box: 
  ${\langle U\rangle = 0}$, ${\psi_{\boldsymbol{n}} = 
  L^{-3/2}\, \mathrm{e}^{i\boldsymbol{p}_{\boldsymbol{n}} 
    \boldsymbol{x}}}$,
${\omega_{\boldsymbol{n}} =   \boldsymbol{p}_{\boldsymbol{n}}^2/2m}$,
$\boldsymbol{p}_{\boldsymbol{n}} = 2\pi  \boldsymbol{n}/L$, and
  ${\boldsymbol{n} \in \mathbb{Z}^3}$. We
assume virialization, i.e.\ statistical independence of different
modes within the gas. Then the correlator of  mode amplitudes
equals,
$$
\langle f_{\boldsymbol{n}} f_{\boldsymbol{n'}}^* \rangle =
f(\boldsymbol{p}_{\boldsymbol{n}}) \delta_{\boldsymbol{nn'}} = 
2\pi^2 \delta_{\boldsymbol{nn'}}\, 
F(\omega_{\boldsymbol{n}})/(m^2 L^3 p_{\boldsymbol{n}})\;,
$$
where the mean phase-space density $f(p_{\boldsymbol{n}}) = \langle
|f_{\boldsymbol{n}}|^2\rangle $ is expressed via~$F\approx
  \langle F\rangle$  which is already time-averaged in
the definition~\eqref{eq:1}. Equation~(\ref{eq:s2}) gives the field
correlator
$$
\langle \psi(t,\, \boldsymbol{x}) \psi^*(t,\, \boldsymbol{y})\rangle =
  \frac{2\pi^2}{m^2 L^3} \int \frac{d^3 \boldsymbol{p}}{(2\pi)^3} \,
  \frac{F(\omega_{\boldsymbol{p}})}{p} \;
  \mathrm{e}^{i\boldsymbol{p}(\boldsymbol{x} - \boldsymbol{y})}\;.
  $$
  Now, we substitute thermal low-$\omega$ asymptotic ${F \to F_0\,
      \omega^{-1/2}}$ at $t\approx t_{gr}$,
    where~$F_0$ is proportional to the effective temperature. At
    large~$|\boldsymbol{x} - \boldsymbol{y}|$ this corresponds
      to a power law,
\begin{equation}
  \label{eq:s4}
  \langle \psi(t_{gr}, \, \boldsymbol{x}) \psi^*(t_{gr},\boldsymbol{y})\rangle
  \approx  \frac{\pi F_0 }{\sqrt{2}\,  m^{3/2} L^3} 
  \; |\boldsymbol{x} - \boldsymbol{y}|^{-1}\,.
\end{equation}
Generically, such power-law behavior is a benchmark of
  second-order critical phenomena. This strongly suggests 
  that Bose star formation is a sister process.

\section*{B. Landau scattering integral}
\label{sec:b.-landau-scattering}
Let us review Landau kinetic equation for the homogeneous and
isotropic gas of gravitating waves~\cite{Levkov:2018kau}. In terms of 
a dimensionless energy distribution
$\tilde{F}(\tau,\,\tilde{\omega})$, it has the form~\eqref{eq:2},  
where 
\begin{equation}
  \label{eq:s1}
  \mathrm{St}\, \tilde{F} = -  \partial_{\tilde{\omega}}
  \tilde{S}(\tau,\, \tilde{\omega})
\end{equation}
is the scattering integral related to the Landau flux~$\tilde{S}$;
hereafter we mark all dimensionless quantities with tildes. The
flux~$\tilde{S}$~--- a cubic functional of $\tilde{F}$ at a
given~$\tau$~---  describes interaction--induced drift of particles in
the phase space:
\begin{equation}
  \label{eq:s3}
  \tilde{S} =  \frac{2^{3/2} b}{3} \, \left\{
  (\tilde{A} - \tilde{B} \tilde{F})\, \frac{\tilde{F}}{2\tilde{\omega}} - \tilde{A}
  \partial_{\tilde{\omega}} \tilde{F} \right\}\,.
\end{equation}
Here~$b$ is the numerical coefficient from~$t_{gr}$, whereas
\begin{align}
  \label{eq:s7}
  & \tilde{A}(\tilde{\omega}) \equiv \int_0^{\infty}
  d\tilde{\omega}'  \, \tilde{F}^2(\tilde{\omega}') \;
  \frac{\mathrm{min}^{3/2}(\tilde{\omega}, 
\tilde{\omega}')}{3 \tilde{\omega}'
    \tilde{\omega}^{1/2}}\,,\\
  \label{eq:s8}
 & \tilde{B}(\tilde{\omega})  \equiv \int_0^{\tilde{\omega}}
d\tilde{\omega}' \tilde{F}(\tilde{\omega}')\,,
\end{align}
see Ref.~\cite{Levkov:2018kau} for derivation and details.

  For us, the most important property of the Landau scattering
  integral is its behavior under the
  scaling~\eqref{eq:ab}. Substituting the latter into 
  Eqs.~(\ref{eq:s3}), (\ref{eq:s7})  and changing integration
  variable to
  $\tilde{\omega}'_s = \beta \tilde{\omega}'$, we find  
    $\tilde{A}(\tilde{\omega}) = \alpha^2 A_s(\beta
    \tilde{\omega})/\beta$ and ${\tilde{B}(\tilde{\omega}) = \alpha
    B_s(\beta \tilde{\omega})/\beta}$, 
  where~$A_s$ and~$B_s$ denote integrals with~$F \to F_s$. We get
  $\tilde{S}(\tilde{\omega}) = \alpha^3 S_s(\beta \tilde{\omega})$
  and  \begin{equation}
    \label{eq:s9}
    \mathrm{St}\, \tilde{F}
  (\tilde{\omega}) = \alpha^3 \beta\, \mathrm{St}\, F_s (\beta
    \tilde{\omega})\,.
  \end{equation}
  This last scaling law is used in Sec.~3 of the main text. 

  Note that the scaling properties of the scattering integral can be 
  understood in a simpler and more general way. To this end we
  partially restore dimensionful units ${F = M \tilde{F}/2\omega_0}$,
  ${\omega = 2 \omega_0 \tilde{\omega}}$, and rewrite kinetic 
  equation~\eqref{eq:2} as ${\partial_t F =  M \,
  \mathrm{St}\, \tilde{F} / (2 \omega_0 t_{gr})}$.  Instead of
  rescaling $\tilde{F}$ and $\tilde{\omega}$ via Eq.~\eqref{eq:ab}, we
  can now change units: 
  $\omega_0 \to \omega_{0}/\beta$ and  $M \to \alpha
  M /\beta$. This gives $t_{gr} \to  t_{gr}/(\alpha^2 \beta)$ and the
  same transformation law of the right-hand side as in Eq.~(\ref{eq:s9}). 
  
\section*{C. Self-similar profiles}
\label{sec:s2.-self-similar}


In the main text, we introduced two modifications of the
  profile equation \eqref{eq:4} to account for condensation. First, we
  impose absorbing boundary condition at ${\omega \approx 0}$:  
  enforce ${F_s = 0}$ at ${\omega_s \leq \omega_{\mathrm{IR}}}$ and
  then send the regulator~$\omega_{\mathrm{IR}}$ to zero. We will see
  that this corresponds to a finite and negative particle flux at
  small~$\omega_s$. Second, we mimic energy income from the
  condensing particles by adding the source~$J_s$ to the right-hand side of
  the equation,
\begin{equation}
  \label{eq:s5}
  (2/D-1) \,\omega_s \partial_{\omega_s}F_s - F_s/D = - 
  \partial_{\omega_s} S_s + J_s(\omega_s)\,.
\end{equation}
Here the scattering integral is expressed via Landau 
  flux~$S_s(\omega_s)$ and the subscript~$s$ means that the flux
  and its sub-integrals $A_s(\omega_s)$, $B_s(\omega_s)$ in
    Eqs.~(\ref{eq:s3})~---~(\ref{eq:s8}) are calculated using $F_s$
    and $\omega_s$ instead of $\tilde{F}$ and~$\tilde{\omega}$.
    
  We turn Eq.~(\ref{eq:s5}) into a set of   
first-order differential equations. First, the
definitions~(\ref{eq:s3})~---~(\ref{eq:s8}) of the scattering  
integrals imply that
$\partial_{\omega_s} A_s =-A_s/2\omega_s  + C_s$ and
$\partial_{\omega_s}  B_s  = F_s$, where $\partial_{\omega_s}C_s =
-F_s^2 / 2\omega_s$. Second, Eqs.~(\ref{eq:s1}) and~(\ref{eq:s5}) can
be viewed as expressions for 
$\partial_{\omega_s} F_s$ and $\partial_{\omega_s}S_s$,
respectively. This totals to five equations for the unknowns $F_s$,
$A_s$, $B_s$, $C_s$, and $S_s$.

The absorbing boundary conditions imply $F_s  = B_s = {3A_s -
  2\omega_{\mathrm{IR}} C_s} = 0$ at $\omega_s  =
\omega_{\mathrm{IR}}$. They leave two Cauchy data
$C_s(\omega_{\mathrm{IR}})$ and $S_s(\omega_{\mathrm{IR}})$ which
serve as shooting parameters. We  tune them to ensure  
regularity: $F_{s},\; C_s \to 0$ as~${\omega_s \to +\infty}$.

At $J_s = 0$ and $D = 5/2$, the profile equation has a scaling
symmetry $F_s\to \alpha_0 F_s(\omega_s/\alpha_0^2)$ with 
arbitrary~$\alpha_0$. This is the case when both conditions at
$\omega_s \to +\infty$ can be satisfied by choosing
$C_s(\omega_{\mathrm{IR}})$, while the flux $S_s(\omega_{\mathrm{IR}})    
\ne 0$ remains unfixed. If the source is nonzero  and ${D > 5/2}$,
the symmetry is  absent, and we obtain one solution per every~$D$
  and~$J_s(\omega_s)$. 

\begin{figure}
  \centerline{\includegraphics{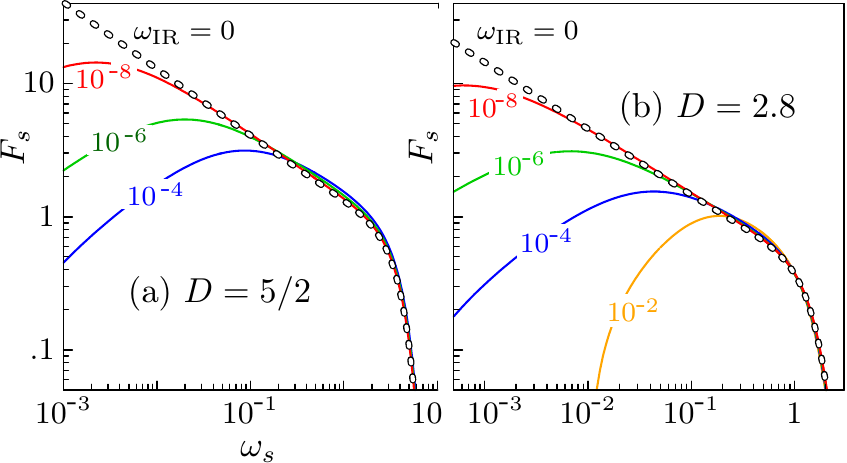}}
  \caption{Self-similar profiles with (a)~$D=5/2$, $J_s = 0$,
    $S_s(\omega_{\mathrm{IR}}) = -1$ and (b)~$D = 2.8$,
    $J_s(\omega_s) = J_0\,\mathrm{sech}^{2}(\omega_s - \omega_1)$,
    $J_0\approx 0.052$, and $\omega_1 = 1.2$. Numbers near the graphs
    give the values  of~$\omega_{\mathrm{IR}}$. \label{fig:self_similar_D2.5}}
\end{figure}

In Fig.~\ref{fig:self_similar_D2.5}(a) we show the solutions
$F_s(\omega_s)$ with 
${D=5/2}$ and $S_s(\omega_{\mathrm{IR}}) = -1$, while
Fig.~\ref{fig:self_similar_D2.5}(b) visualizes the case $J_s \ne 0$
and ${D=2.8}$.  It is clear that the self-similar profiles have definite 
limits~$\omega_{\mathrm{IR}} \to 0$. Indeed, Eq.~(\ref{eq:s5})
suggests an infrared asymptotic~\footnote{This solution 
  satisfies $C_s(\omega_s) - F_{0s}^2/2\omega_s \to 0$ as $\omega_s
  \to 0$.}
\begin{equation}
  \label{eq:s6}
  F_s =  F_{0s}\,  \omega_s^{-1/2} + F_{1s}\, \omega_s +
  O(\omega_s^{3/2})   \dots \;\; \mbox{as} \;\; \omega_s \to 0
\end{equation}
in the unregularized case $\omega_{\mathrm{IR}}=0$, where
$F_{0s}$, $F_{1s}$ are constants. Imposing this behavior, we obtain the
${\omega_{\mathrm{IR}} = 0}$ graphs in Fig.~\ref{fig:self_similar_D2.5}
(chain points). Note that  Eq.~\eqref{eq:s6} includes a thermal
  tail at $\omega_s \to 0$, which is indeed observed at
$\omega_{\mathrm{IR}}  \ll \omega_s \ll \omega_1$ in the full
numerical simulations, see Fig.~\ref{fig:spectra}(c)  from the main
text.

The profile with $\omega_{\mathrm{IR}} = 10^{-2}$ from
  Fig.~\ref{fig:self_similar_D2.5}(b) is repeated in
  Figs.~\ref{fig:spectra}(a) and~(c) of the main text. It has 
  $J_s(\omega_s) = J_0\,\mathrm{sech}^{2}(\omega_s -
  \omega_1)$, $J_0\approx 0.052$, and $\omega_1 = 1.2$. These 
  parameters are selected to fit the simulation data.

Another good remark is that the self-similar profiles fall off as fast
as ${F_s \propto   \omega_s^{q}\, \mathrm{e}^{- \zeta \,
      \omega_s^{5/2}}}$ at ${\omega_s \to +\infty}$, where $q =
    (4-D)/(2D-4)$ and $\zeta = (2D-4) / \lim\limits_{\omega_s\to \infty}
  {(5DA\sqrt{\omega_s}) }$. The 
  cutoff appears because gravitational scattering is ineffective at high
  $\omega$. Above the cutoff, the particles cannot  participate in
  self-similar dynamics.

To summarize, the profile equation~(\ref{eq:s5}) has two families
  of nontrivial solutions: one solution per every $J_s$ at $D  >
  5/2$ and a branch of $D = 5/2$ solutions with arbitrary condensation
  flux $S_s(\omega_{\mathrm{IR}})$ and zero source.

\section*{D. Attracting to self-similar solutions}
\label{sec:s1.-additional-data}

\begin{figure}
  \unitlength=1mm
  \centerline{\begin{picture}(86, 52)
    \put(0,0){\includegraphics{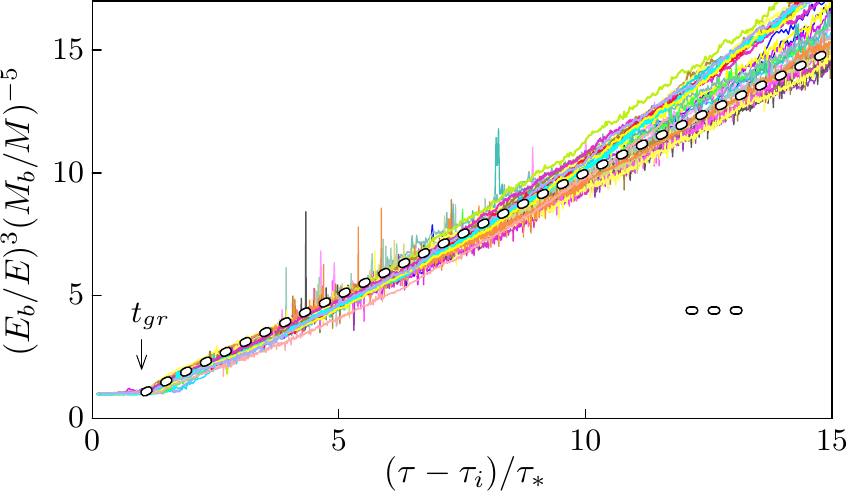}}
    \put(57,17.5){Eq.~\eqref{eq:result}}
  \end{picture}}
  \caption{Evolutions of the ratio $E_b^3/M_b^{5}$ in all our
    Schr\"odinger-Poisson simulations (pale solid lines). These
      runs have essentially different $t_{gr}$, see
    Sec.~4 of the main text. Chain points show the
    law~\eqref{eq:result} with ${\tau_i 
      =  -0.1}$ and ${\tau_* = 1.1 \cdot (1 - x_e)^5}$.}  
  \label{fig:SP8}
\end{figure}

\begin{figure}[b!]
  \centering
  \includegraphics{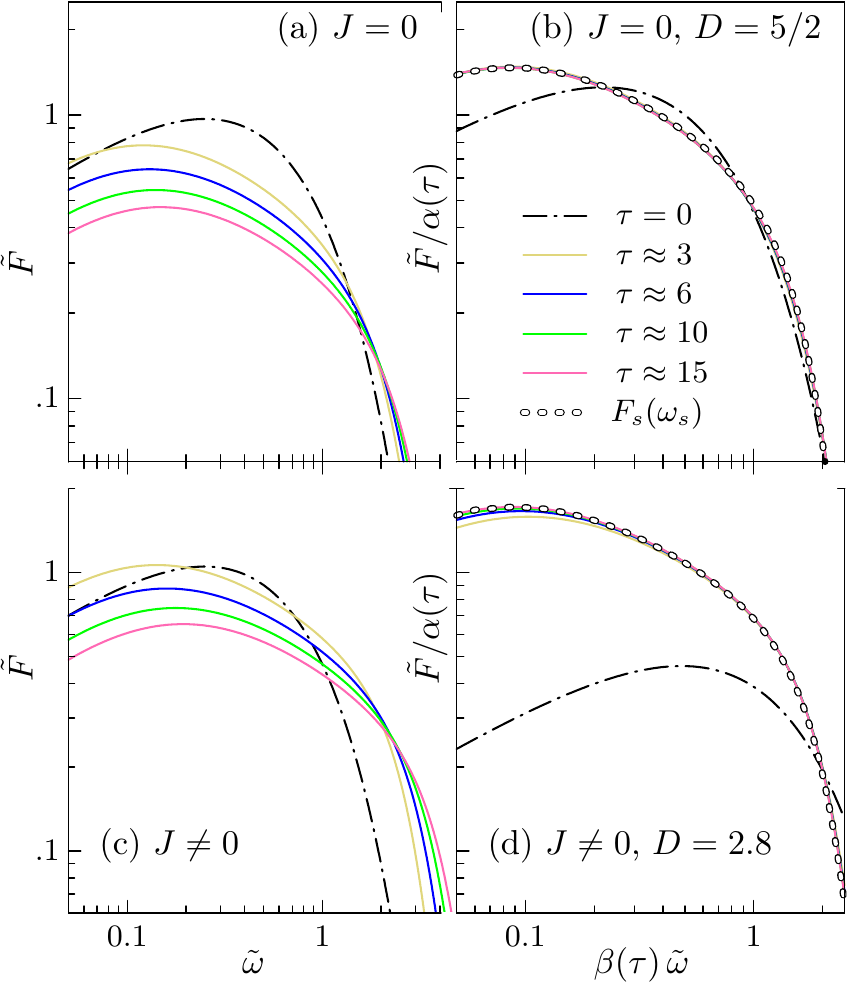}

  \includegraphics{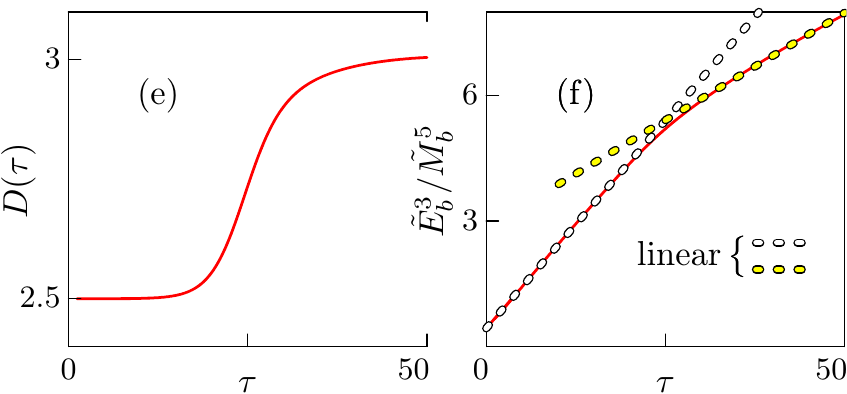}
 
  \caption{Numerical solutions of the modified kinetic
    equation~\eqref{eq:S21}. Figure~(a) is plotted for
    $\tilde{J}=0$, while (c)~considers self-similar source $\tilde{J}=
    \alpha(\tau) J_s(\beta(\tau)\tilde{\omega})/(\tau-\tau_i)$ with $J_s(\omega_s) = 
    J_0  \, \mathrm{sech}^{2}(\omega_s - \omega_1)$,  $J_0\approx 0.1$, $\omega_1
    =  1.2$,  $D=2.8$, and $\tau_i  = -0.1$. (b),~(d)~Transformations
    \eqref{eq:ab}, (\ref{eq:S22}) of the spectra~(a) and (c)  with parameters
    (b)~$D=5/2$, $\tau_i \approx -1.9$ and (d)~$D = 2.8$, ${\tau_i =
      -0.1}$ (lines). Chain points show self-similar profiles $F_s(\omega_s)$
    with regulators (b)~$\omega_{\mathrm{IR}} =  2\cdot    10^{-3}$ and
    (d)~$\omega_{\mathrm{IR}} = 4.5\cdot 10^{-4}$.
     (e), (f)~Evolutions of ${D = D(\tau)}$ and $E_b^3/M_b^5$ in the
      case of essentially non self-similar source (lines). }
    \label{fig:kin}
\end{figure}

Let us demonstrate that the self-similar solutions \eqref{eq:ab} are 
attractors of kinetic evolution.

To warm up, we explicitly test Eq.~\eqref{eq:result} using
  full Schr\"odinger-Poisson simulations. Recall that this law  
  approximately describes self-similar bath with slowly-varying $D =
  D(\tau)$. We compute the bath masses $M_b(t)$ and energies 
$E_b(t)$ for all our solutions from 
    Figs.~\ref{fig:spectra} and~\ref{fig:Mbs} using the 
distribution functions $F(t,\,\omega)$ at~${\omega>0}$.  This gives 
  $M_b = M  - M_*$ and $E_b = E -  E_*$. Then in
  Fig.~\ref{fig:SP8} we plot  the left-hand side of
Eq.~\eqref{eq:result} versus the right-hand 
side, i.e.\ basically $E_b^3/M_b^5$ as functions of~$\tau$
(pale solid lines). 
The resulting curves are close to each other and are well 
described by Eq.~\eqref{eq:result}  
(chain points)  despite essentially different parameters of the
solutions. This confirms self-similar character of  the bath
evolution  and, hence, our theory for Bose star growth. 

Next, we consider time-dependent kinetic
  equation~\eqref{eq:2},~(\ref{eq:s1}). To account for condensation
  onto the Bose star, we introduce an absorbing sink $\tilde{W}$ at
  $\tilde{\omega} \approx 0$ and an energy source~$\tilde{J}$,
\begin{equation}
  \label{eq:S21}
  \partial_\tau \tilde{F} = -\partial_{\tilde{\omega}}
  \tilde{S} - \tilde{W}(\tilde{\omega}) \tilde{F} + \tilde{J}(\tau,\,
  \tilde{\omega})\,.
\end{equation}
In practice, we use the sink profile $\tilde{W}= W_0\,
\mathrm{e}^{-(\tilde{\omega}  / \omega_{\mathrm{IR}}')^2}$  
concentrated at $\tilde{\omega} \lesssim
\omega_{\mathrm{IR}}'$. It effectively destroys low-energy
  particles at   $W_0 =   200/\omega_{\mathrm{IR}}'$ and 
  $\omega_{\mathrm{IR}}' = 4\cdot 
  \mathrm{10}^{-4}$. We start simulations from the 
Gaussian-distributed (virialized) initial state ${\tilde{F}(0,\,
  \tilde{\omega})  \propto \tilde{\omega}^{1/2}\,
  \mathrm{e}^{-2\tilde{\omega}}}$ and play with
different~$\tilde{J}(\tau,\, \tilde{\omega})$.

Figure \ref{fig:kin}(a) shows the numerical solution
$\tilde{F}(\tau,\, \tilde{\omega})$ at ${\tilde{J} = 0}$. This is the
case when the energy of the bath is (almost)
conserved and
the mass is not: recall that the sink swallows particles with
$\tilde{\omega} \approx  0$.  The self-similar profile with
  such properties has $D\approx 5/2$, see
  Eq.~\eqref{eq:k}. In Fig.~\ref{fig:kin}(b) (dash-dotted and solid
  lines) we perform self-similar rescaling of the
  spectra~(a) with $D = 5/2$ and
  \begin{equation} 
  \label{eq:S22}
  \alpha(\tau) = (\tau-\tau_i)^{-1/D}\,, \;\;\;  \;\;\;
  \beta(\tau) = (\tau-\tau_i)^{2/D -1}\,,
\end{equation}
where the time-translation parameter $\tau_i$ is   restored as
compared to Eq.~\eqref{eq:ab}.  We see that for properly adjusted $\tau_i\approx
-1.9$ all of the rescaled graphs except for the one with
$\tau=0$ merge into a single curve coinciding with 
$D=5/2$ self-similar profile $F_s(\omega_s)$ (chain points). It is
worth noting that the 
absorbing sink is implemented differently in our calculations of $F_s$
and $\tilde{F}$~--- hence the difference
in their infrared regulators $\omega_{\mathrm{IR}}$ and 
$\omega_{\mathrm{IR}}'$. We see 
 that although the starting distribution does not resemble the
self-similar profile at all, the evolved spectra
  approach $\alpha F_s(\beta \tilde{\omega})$  at $\tau \sim
3$ and remain close to it at later times. This 
  proves that the self-similar 
  solution with $D = 5/2$ is an attractor at $\tilde{J} = 0$. 

Now, add the energy source with self-similar time dependence:
$\tilde{J} = \alpha(\tau)  J_s(\beta(\tau)
\tilde{\omega})/(\tau-\tau_i)$, where $J_s(\omega_s)$ has the same 
form as in Fig.~\ref{fig:spectra};  $D = 2.8$ and $\tau_i = -0.1$. The
respective solution $\tilde{F}(\tau,\, \tilde{\omega})$ of the
 kinetic 
equation is visualized in  Fig.~\ref{fig:kin}(c). At late times, it exhibits
the self-similar behavior with $D =
2.8$. Indeed,  the rescaled spectra in Fig.~\ref{fig:kin}(d) (lines)
coincide at $\tau > 3$ with  the $D=2.8$ self-similar profile 
(chain points). Again, we see that the self-similar solutions are
attractors. 

Note that the scaling weight $D$ of the solution does not always
correspond to the time dependence of the external source. If the  
amplitude of $\tilde{J}$ is too large or too small, the function $\tilde{F}(\tau,\, 
\tilde{\omega})$ first attracts to the self-similar profile with
different $D$. Later, the weight starts to evolve slowly until the
source-prescribed value is reached. In such a case, the
dynamics remains approximately self-similar at all times but $D$
slowly drifts with~$\tau$.  

The latter situation is illustrated in Figs.~\ref{fig:kin}(e),~(f),
  where we consider the source $\tilde{J} = J_0\,
  \vartheta(\tau) \, \mathrm{sech}^2(\tilde{\omega} -  \omega_1)$
  switching on at $\tau \sim 25$ as $\vartheta(\tau) =  [1 +
    \mathrm{e}^{(\tau_i' - \tau)/\Delta \tau}]^{-1}(\tau  - 
  \tau_i)^{-2/3}$, where $J_0 \approx 0.017$, 
$\omega_1 = 1.2$, $\tau_i \approx -0.6$, 
  $\tau_i' \approx 24$, and $\Delta \tau \approx 2.3$. This 
   time dependence of $\tilde{J}(\tau,\, \tilde{\omega})$
  explicitly breaks the scaling symmetry of 
  Eq.~(\ref{eq:S21}) making the parameter $D  =(3R-5)/(R-2)$ in
  Fig.~\ref{fig:kin}(e) jump from $D \approx 5/2$ in the beginning of
  the process to almost $3$ in the end; to plot the figure, we
    extracted $R \equiv d  \ln   
E_b / d\ln M_b$ from the numerical
evolution. Nonetheless, the combination $E_b^3/M_b^5$ [solid line in
  Fig.~\ref{fig:kin}(f)] becomes almost linear in the late-time
region where $D = D(\tau)$ starts to evolve slowly, again~--- see the
linear fits (chain points). This confirms that the solution attracts to
self-similarity even after the strong kick at $\tau\approx   25$.

It is
worth noting, however, that the tilts of the linear graphs in   Fig.~\ref{fig:kin}(f)
are different at early and late times. This 
  implies that Eq.~\eqref{eq:result} holds, but the parameter $\tau_*$
  insubstantially changes with time. The latter change is ignored in the main
  text but should be taken into account in the refined approaches.
  
To summarize, we numerically proved that self-similar
solutions~\eqref{eq:ab} are attractors of kinetic evolution
with a sink at $\tilde{\omega} \approx 0$ and an
  energy source.

\section*{E. Self-interacting bosons}
\label{sec:d.-self-interacting}

\begin{figure}

  \vspace{3mm}
   \centerline{\includegraphics{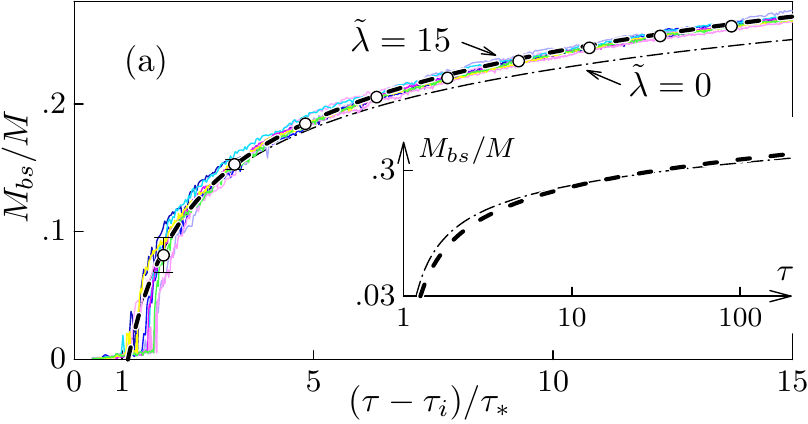}}

   \vspace{1mm}
   \centerline{\includegraphics{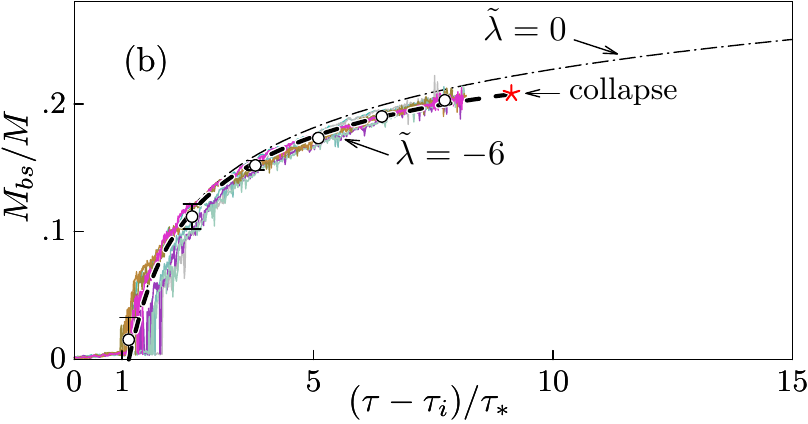}}
   \caption{Mass evolution of Bose stars in the model with nonzero
     self-coupling at $\epsilon \approx 0.186$ and
     (a)~$\tilde{\lambda} = 15$, (b)~${\tilde{\lambda} = -6}$. Thin
     color lines display results of 9+8 simulations, whereas
     thick dashed is the theory with (a)~${\tau_i \approx -0.51}$, $x_e
       \approx 0.023$; (b)~$\tau_i \approx 0.14$, $x_e \approx
       0.026$. For reference, we repeat the
       theoretical curve with~$\lambda=0$ and $\epsilon \approx
       0.186$ from Fig.~\ref{fig:Mbs}(b) (thin dash-dotted line). The
       inset of Fig.~(a) shows the theoretical curves with
       $\tilde{\lambda} = 0$ and~$\tilde{\lambda} = 15$ at larger
       timescales.
   }
  \label{fig:self-int}
\end{figure}

A highly nontrivial test~\footnote{We thank the Referee for
  suggesting this check.} of our theoretical framework can be
performed by studying growth of Bose stars in the bath of
self--interacting bosons. We describe self-interactions by
adding the term~$\lambda |\psi^2|\psi / (8m^2)$ with coupling 
constant $\lambda$ to the right--hand side of the upper
Eq.~\eqref{eq:GP2}. This upgrades the full set to
Gross--Pitaevskii--Poisson system.   

  In the presence of gravity, a comparative effect of
  self-interactions is characterized by a dimensionless
  combination~\cite{Dmitriev:2021utv} $\tilde{\lambda} =  2\lambda
  \omega_0 / (m^3 G)$, where~$\omega_0$ is the typical particle
  energy. Our self-similar solution~\eqref{eq:result}
  is applicable if gravity dominates,
  i.e. at~\cite{Levkov:2018kau, Chen:2021oot}
  \begin{equation}
    \label{eq:S23}
    \frac{\tau_{gr}}{\tau_{\lambda}} \sim\ \frac{\sigma_{\lambda}}{\sigma_{gr}} \simeq
    \frac{\tilde{\lambda}^2}{1024 \pi^2\ln(p_0L)} \ll 1\,.
  \end{equation}
  Here $\tau_{gr}$ and $\tau_\lambda$ are the gravitational and
  self-interaction relaxation times, while $\sigma_{gr}$
  and~$\sigma_{\lambda}$ are the respective transport cross
  sections. Below we keep~\footnote{This ratio is even smaller in
    magistral cosmological models. For example, $\tau_{gr}/\tau_{\lambda} \sim 
    10^{-12}$ and $\tilde{\lambda} \sim 10^{-4}$ \cite{Levkov:2018kau,
      Chen:2021oot} inside QCD axion miniclusters. At these
    values, the effect of self-interactions on growth of Bose
      stars is negligible. To make them relevant, one 
    switches~\cite{Chen:2020cef, Chen:2021oot} to general  
    ``axion--like'' models with deliberately 
    enlarged~$\lambda$.}  $\tau_{gr}/\tau_{\lambda} \lesssim 10^{-2}$ 
  in  all simulations. 

\begin{figure}[t]
  \centerline{\includegraphics{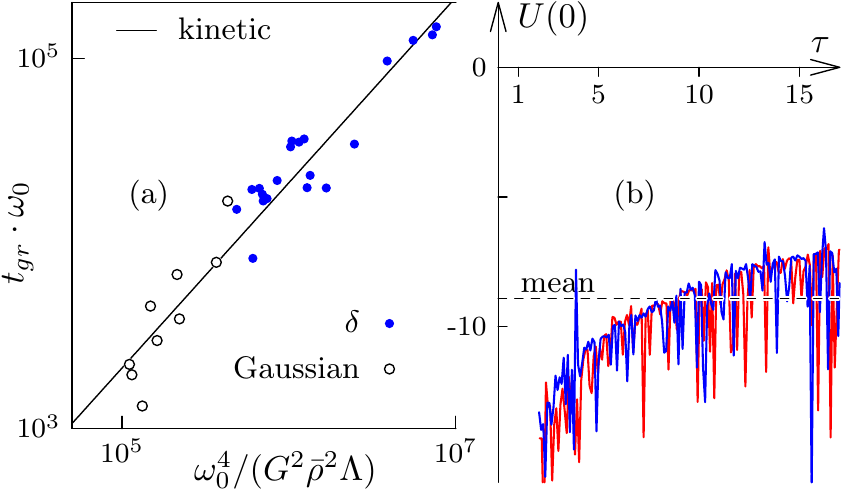}}
  \caption{(a)~The times $\omega_0 t_{gr}$ of Bose stars formation
      in miniclusters versus the relevant factor in the theoretical expression
      for this quantity; $\Lambda  \equiv \ln(p_0 R)$ is the Coulomb
      logarithm and $R$ is the minicluster size. Filled and empty
      points correspond to simulations of Ref.~\cite{Levkov:2018kau}
      starting from the~$\delta$--distributed gas and our solutions
      with Gaussian initial conditions, respectively. Thin solid line  
      is the theory with~$b = 0.7$. (b)~Time--dependent 
      potential~$U(0)$ in the minicluster center (dimensionless
      units). Thin solid graphs are extracted from two long
      simulations, while the {dashed} line is the time--averaged value.
      \label{fig:mc_tests}}    
\end{figure}

  Despite Eq.~\eqref{eq:S23}, self--interactions can modify the growth
  law of Bose stars, and their effect can be
  considerable~\cite{Chen:2020cef}. Indeed, although the terms
    proportional to $\lambda$ are small inside the light stars,
  they grow with mass and start to 
  dominate~\cite{Chavanis:2011zm} at $M_{bs} \gtrsim M_{\lambda}
  \equiv |\lambda G|^{-1/2}$. If the self-interactions are
  repulsive,~$\lambda > 0$, this increases the energy of heavy stars
  to $E_{bs} \propto -M_{bs}^2$. The attractive case is more dramatic:
  negative self-pressure makes the Bose stars collapse~\cite{Chavanis:2011zm,
    Eby:2016cnq, Levkov:2016rkk, Chen:2020cef} as Bosenovas at $M_{bs}
  \gtrsim 10.2 \, M_{\lambda}$. Generically, we write 
  \begin{equation}
    \label{eq:S24}
    E_{bs} = - 
    \gamma M_{bs}^3 \, {\cal E}(M_{bs}/M_\lambda)\,,
  \end{equation}
  where the function ${\cal E}$ accounts for self-interaction
  energy. It is smaller (larger) than $1$ at $\lambda > 0$ ($\lambda <
  0$). Besides, ${\cal E} \approx 1$ at $M_{bs} \ll M_{\lambda}$ when
  the self-interactions are negligible. In practice,   we compute ${\cal 
    E}$ numerically by solving the Gross--Pitaevskii--Poisson system
  for every~$M_{bs}/M_{\lambda}$.

  Using Eq.~(\ref{eq:result}), we obtain the growth law of
  self-interacting stars [cf.\ Eq.~\eqref{eq:3}], 
  \begin{equation}
  \label{eq:s26}
  (1 + x_{bs}^3{\cal E}/\epsilon^2)^3 (1 - x_e
  - x_{bs})^{-5} \approx (\tau - \tau_i) / \tau_*\,,
  \end{equation}
  where ${\cal E} \equiv {\cal E}(x_{bs}M/M_{\lambda})$. At the
  qualitative level, Eq.~(\ref{eq:s26})  
  agrees with the phenomenon suggested in Ref.~\cite{Chen:2020cef}:
  for fixed $\tau_*$ and~$\tau_i$ the Bose stars grow faster for
  positive~$\lambda$ (${\cal E}<1$) and slower for
  negative~$\lambda$~(${\cal E}<1$). This feature is illustrated in 
  Figs.~\ref{fig:self-int}(a), (b) that compare two theoretical
  curves~$x_{bs}(\tau)$, Eq.~(\ref{eq:s26}), at $\tilde{\lambda}
  = 15$ and~$\tilde{\lambda} = -6$  (thick dashed lines) with the one at
  $\lambda=0$ (thin dash-dotted).  

  We performed an explicit numerical test of
  Eq.~(\ref{eq:s26}). Starting from the Gaussian-distributed gas in
  the box, we performed~9 simulations at $\tilde{\lambda}=15$ and~8
  simulations at~$\tilde{\lambda}=-6$. We used $M = 20 \, p_0/m^2 G$
  and ${L = (35 \div  40)/p_0}$. Mass evolutions of the
    respective Bose stars are shown in
  Figs.~\ref{fig:self-int}a,~b by thin color lines. They are well
  described by Eq.~(\ref{eq:s26}) (thick dashed). However, the
  values of the fitting parameter~\footnote{We still extract $x_e$ from
    the distribution function and compute~${\epsilon^2 \equiv E/\gamma
      M^3}$ using the initial data. The value of
    $\tau_*$ is fixed by the condition $M_{bs}=0$ at $\tau=1$, see Sec.~4.} 
  $\tau_i$ are different with respect to non self--interacting case: we
  obtain $\tau_i \approx -0.51$ at $\tilde{\lambda} = 15$ and ~$\tau_i
  \approx 0.14$ at~$ \tilde{\lambda} = -6$. 

  It is worth noting that the dependence of $\tau_i$  on  $\tilde{\lambda}$ 
  affects the growth law of Bose stars. 
  At moderately small~$\tau$, it may even compensate the 
  (de)acceleration effect of self--interaction energy, see\ the inset in
  Fig.~\ref{fig:self-int}(a). However, at large timescales the
  self--energy wins and makes the growth go faster at $\lambda > 
  0$ and slower at $\lambda < 0$~--- see the inset, again. 

\section*{F. Simulations in miniclusters}
\label{sec:e.-simul-minicl}
Although the application of our theory~\eqref{eq:3} is straightforward
  at the qualitative level, things become more tricky once precise
  agreement with simulations is required. To this end, we accurately
  determine the minicluster parameters. 

  We form gravitationally bound miniclusters by triggering strong
  Jeans instability in the dense virialized gas~\cite{Levkov:2018kau,
    Chen:2020cef}. In particular, our two long simulations start from
  very large mass $M_{\mathrm{tot}} = 112.5/\omega_0$ in the box~$L =
  52.5/p_0$. At these values, the miniclusters engulf more than $55\%$
  of matter, and the remaining diffuse particles do not affect much the
  growth of objects within them. 

  We define the minicluster center as the center-of-mass of matter
  distribution within the box; we call it $\boldsymbol{x}=0$ for
  simplicity. The density $\bar{\rho} = \rho(0)$ in the minicluster
  center is then obtained as the value of $\rho =
  m|\psi(t,\, \boldsymbol{x})|^2$ averaged over the Gaussian spatial
  window. The remaining parameters are extracted from the distribution 
  function~\eqref{eq:1} or, specifically, from its part at $\omega <
  0$ that describes a self-bound minicluster. Namely, the mass $M$ and
  energy $E_{\mathrm{mc}}<0$ of the minicluster are obtained by
  integrating $F$ and $\omega F/m$ over this region. Then the
    virial particle energy equals
  $\omega_0 \equiv -mE_{\mathrm{mc}}/M$, the virial radius
    is $R = (3\omega_0/2\pi m G\bar{\rho})^{1/2}$, while $\Lambda
    = \ln (p_0 R)$ is the Coulomb logarithm.

  Once the minicluster parameters are specified, we find 
  the numerical factor~\footnote{This is a necessary part of the
    procedure compensating for our voluntary choice of the 
    minicluster parameters $\bar{\rho}$, $\omega_0$, and $R$.}~$b$ in
  the expression for the relaxation time~$t_{gr}$. To this end we
  perform many short-time simulations at different values of
  parameters and wait until Bose stars appear in their miniclusters. The
  moments $t_{gr}$ when they form (empty points 
  in Fig.~\ref{fig:mc_tests}(a)) are well described by the theory with~$b
  = 0.7$ (line)~--- the same value as in Ref.~\cite{Levkov:2018kau}. The
  coincidence of $b$'s is remarkable because minicluster simulations of
  Ref.~\cite{Levkov:2018kau} (filled points) start from 
  the $\delta$-distributed gas in the box, $|\psi_{\boldsymbol{p}}|^2 \propto 
  \delta(|\boldsymbol{p}| - p_0)$, while our simulations use
  Gaussian gas with $|\psi_{\boldsymbol{p}}|^2 \propto 
  \mathrm{e}^{-\boldsymbol{p}^2/p_0^2}$. This suggests that 
  formation of miniclusters strongly intermixes
  the gas forcing it to ``forget'' the initial condition.
  
  An important part of our procedure is a computation of
  the gravitational potential $U(0)$ in the minicluster center. In
  the notations of Eqs.~\eqref{eq:result}, \eqref{eq:3}, this
    parameter enters the 
  total energy $E = E_{\mathrm{mc}} - U(0) M$ which is positive  and
  counted from the lowest level inside the minicluster. Notably, the
  value of $U(0)$ visibly drifts with time, since the minicluster gets
  eaten by the Bose star and becomes lighter; see
  Fig.~\ref{fig:mc_tests}(b). We calculate the potential using the Bose
  star itself as a sensor. On the one hand, its mass $M_{bs}$ and
  binding energy $\omega_{bs} = -3m\gamma M_{bs}^2$ can be extracted
  from the profile $|\psi_{bs}(\boldsymbol{x})|^2$. On the other, the
  ``Bose star'' peak in the energy distribution is located at $\omega
  = \omega_{bs} + mU(0)$. Subtracting these quantities, we obtain the
  solid lines in Fig.~\ref{fig:mc_tests}(b) corresponding to two long simulations.
  We use the time-averaged value of $U(0)$ (dashed
  horizontal line) in the theoretical expressions for $E$ and
  $\epsilon^2 = E/\gamma M^3$.

  Finally, we determine the fraction $x_{e} = M_e/M$ of particles
  on the discrete levels of the Bose star potential in
  the same way as before: by integrating $F$ over the region $\omega_{bs} + mU(0) <
  \omega < mU(0)$.  Once this is done, the theoretical predictions
  (\ref{eq:s26}) match the Bose star mass curves extracted
    from the simulations (lower graph in
  Fig.~\ref{fig:Mbs}(b)). The respective best-fit value $\tau_i
    \approx -0.1$ matches that in the box simulations.


\bibliographystyle{aipnum4-1}
\bibliography{refs}

\end{document}